\documentstyle[preprint,aps,prb]{revtex}
\tightenlines
\begin{document}
\draft

\title{Phonon drag thermopower and weak localization}

\author{A. Miele, R. Fletcher and E. Zaremba}
\address{Physics Department, Queen's University, Kingston, Ontario,
Canada, K7L 3N6.}

\author{Y. Feng}
\address{Microstructural Sciences, National Research Council, Ottawa,
Ontario, Canada, K1A 0R6.}

\author{C. T. Foxon}
\address{Department of Physics, University of Nottingham,
Nottingham NG7 2RD, UK.}

\author{J.J. Harris}
\address{Electronic and Electrical Engineering, University College,
London WC1E 7JE, UK.}

\date{7 July, 1998}
\maketitle

\begin{abstract}
Previous experimental work on a two-dimensional (2D) electron gas in a
Si-on-sapphire device led to the conclusion that both conductivity
and phonon drag thermopower $S^g$ are affected to the same
relative extent by weak localization. The present paper presents
further experimental and theoretical results on these
transport coefficients for two very low mobility 2D electron
gases in $\delta-$doped GaAs/Ga$_x$Al$_{1-x}$As quantum wells.
The experiments were carried out in the temperature range 3-7K where
phonon drag dominates the thermopower and, contrary to the
previous work, the changes observed in the thermopower
due to weak localization were found to be an order of
magnitude less than those in the conductivity.
A theoretical framework for phonon drag thermopower in 2D and 3D
semiconductors is presented which accounts for this insensitivity
of $S^g$ to weak localization. It also provides transparent
physical explanations of many previous experimental and theoretical
results.
\end{abstract}

\section{Introduction}
\label{introduction}

Weak localization (WL) refers to the quantum correction to the conductivity
which arises because the phase of the electronic wavefunction
is not randomized during elastic collisions. Through interference effects,
this leads to a decrease in the longitudinal conductivity $\sigma_{xx}$
which can be restored either by raising the temperature,
thus increasing the frequency of phase-disrupting inelastic collisions,
or by applying a low magnetic field which destroys the phase coherence
of the electronic paths. WL has been thoroughly
investigated\cite{wlreviews} over the last two decades but the effect
on other transport coefficients has received much
less attention. It was predicted\cite{Castellani87}
that the electrical conductivity $\sigma_{xx}$
and the electronic thermal conductivity
$\lambda^e_{xx}$ would be related by the Wiedemann-Franz relation in
the usual way for elastic electronic scattering, and the
available experimental data\cite{Bayot90} on the longitudinal
components in a quasi-2D system are in accord with this prediction.
(We take the 2D electron gas (2DEG) to be in the $xy$ plane
and, unless otherwise stated, the magnetic field ${\bf B}$ along $z$).

The interpretation of thermopower data is
complicated by the fact that there are two contributions,
diffusion $\mathop{S^d}\limits^\leftrightarrow$
and phonon drag $\mathop{S^g}\limits^\leftrightarrow$.
There is theoretical consensus that the
longitudinal diffusion thermopower $S^d_{xx}$
should be modified by WL,\cite{Afonin84,Castellani88,KandB88,Hsu89}
the result being equivalent to the well-known Mott relation
between $S^d_{xx}$ and $\sigma_{xx}$ for elastic electronic
scattering. However, in 2DEGs, which are the focus of the
present paper, $S^g_{xx}$ is very large at low
temperatures so that $S^d_{xx}$ is difficult to observe.
This situation is quite different from that encountered in 3D
systems. Introducing impurities into a 3D lattice to reduce
the elastic mean free path $l_e$ of the electrons (so that WL is
observable in $\sigma_{xx}$) inevitably leads to a strong
reduction of phonon drag because of the scattering of phonons
by the same impurities. In the 2D case, $l_e$ is reduced by
placing impurities into or near the 2DEG,
but under normal circumstances these impurities have no effect on
the phonons in the substrate and so drag is not thereby reduced.

As far as we aware, there is only one theoretical paper\cite{Afonin92}
concerned with WL and phonon drag. This dealt with $S^g_{xx}$ but gave
a result only for the range of magnetic field $B$ where the
localization term in $\sigma_{xx}$ is quadratic in $B$, which
is appropriate only at very low $B$, and the situation at arbitrary
fields remained unclear.

Experimental data for a 2DEG in a Si-on-sapphire device were published
by Syme {\em et al.}\cite{Syme1and2} which indicated
that $S^g_{xx}$ and $\sigma_{xx}$ are affected equally by WL.
The significance of this result is most conveniently discussed, as
did the authors of that work,
in terms of the phenomenological equation relating the
electric current density ${\bf J}$ to the gradient of the
electrochemical potential ${\bf E^\prime = E} + \nabla \mu /e$, where
${\bf E}$ is the electric field and $\mu$ the chemical potential,
and the temperature gradient $\nabla T$, i.e.
\begin{equation}
\label{Ohm}
{\bf J} = \mathop{\sigma}\limits^{\leftrightarrow}
{\bf E^\prime} -\mathop{\epsilon}\limits^{\leftrightarrow}
\nabla T
\end{equation}
where $\mathop{\epsilon}\limits^{\leftrightarrow}$ is
 a thermoelectric tensor. The thermopower
 $\mathop{S}\limits^{\leftrightarrow}$ is the experimental quantity
which gives access
to $\mathop {\epsilon}\limits^{\leftrightarrow}$ and is defined
by ${\bf E} =  \mathop{S}\limits^{\leftrightarrow}\nabla T$
(the chemical part $\nabla \mu /e$ is not observed in experimental
situations, e.g. see Gurevich {\it et al.}\cite{Gurevich95})
measured under the condition ${\bf J} =0$.
The coefficient $\mathop{\epsilon}\limits^{\leftrightarrow}$
has two additive contributions, diffusion
$\mathop{\epsilon^d}\limits^{\leftrightarrow}$
and phonon drag $\mathop{\epsilon^g}\limits^\leftrightarrow$,
which are responsible for the two corresponding components
$\mathop{S^d}\limits^\leftrightarrow$ and
$\mathop{S^g}\limits^\leftrightarrow$.

$S^g_{xx}$ varies rapidly with temperature and it is difficult to
observe deviations that might be due to
localization. It is therefore necessary to measure $S_{xx}$
as a function of perpendicular magnetic field ${\bf B}$.
From Eq.~(\ref{Ohm}) with $\nabla T_y =0$
we have $S_{xx} = \rho_{xx}\epsilon_{xx} + \rho_{yx}\epsilon_{xy}$,
where $\mathop{\rho}\limits^\leftrightarrow =
{\mathop{\sigma}\limits^\leftrightarrow} ^{-1}$.
The very low mobility of the previous (and present) samples means that
it is an excellent approximation to write $S_{xx} = \epsilon_{xx}/
\sigma_{xx}$. If we denote the relatively small changes due
to $B$ as $\triangle\sigma_{xx}$, etc., then we have
\begin{equation}
\frac{\triangle S^g_{xx}}{S^g_{xx}}
= \frac{\triangle\epsilon^g_{xx}}{\epsilon^g_{xx}}
- \frac{\triangle\sigma_{xx}}{\sigma_{xx}}
\end{equation}
and the data were consistent with the first term on the right hand
side being zero. Indeed in a further publication\cite{Syme3}
it was assumed that this was precisely so and residual field
dependent effects at the level of $< 1\%$ were analyzed in terms
of WL corrections to $S^d_{xx}$.

For reasons which will be explained in detail in Section \ref{theory},
this experimental result is unexpected. Basically we will show that
$\epsilon^g_{xx}$ is proportional to the electron-impurity scattering
rate, $1/\tau_{ei}$, whereas $S_{xx}$ is not.
One can view WL to be a modification of $\tau_{ei}$
due to interference effects. Thus, we would have expected both
$\epsilon_{xx}^g$ and $\sigma_{xx}$ to be affected by WL, but not
$S^g_{xx}$. This is just the opposite of that found by
Syme {\it et al.}\cite{Syme1and2}

In view of these considerations, we have reexamined the question
both experimentally and theoretically.
The present paper gives new experimental data on two samples
of 2DEGs in $\delta$-doped GaAs/Ga$_{1-x}$Al$_x$As quantum wells.
In contrast to the previous results,
we find $\triangle S^g_{xx} \approx 0$ for $B \le 1$~T, which implies
$\triangle \sigma_{xx}/\sigma_{xx}  =
\triangle\epsilon^g_{xx}/\epsilon^g_{xx}$
(to an accuracy of  $\sim 10\%$).

For several years we have been using various physical pictures of
phonon drag thermopower to help in understanding experimental
data on 2D and 3D systems in a qualitative way, e.g.
Refs.~\onlinecite{HGSE} and \onlinecite{Tieke}.
In Section~\ref{theory},
we outline a theory of phonon drag thermopower which puts these ideas
on a quantitative basis. In particular, it provides arguments
to support the result that WL has no influence on phonon drag.
However, we also show that the theory provides a useful framework
to understand many previously published experimental
and theoretical results in a transparent physical way. A brief,
preliminary account of some of this work has been presented in
conference form.\cite{icps}

\section{Experimental Techniques and Results}
\label{expts}
Both our samples have been well characterized in previous
work.\cite{samples} The electron densities are $n=3.6\times
10^{16}$\,m$^{-2}$ for sample 1 (S1) and $n=2.21\times 10^{16}$\,m$^{-2}$
for sample 2 (S2). Using the 4\,K values of the conductivities,
each has a mobility of $\mu \approx 0.12$\,m$^2$/Vs.
Only a single subband is occupied in each sample.

The thermoelectric properties of S1 have been previously studied
as a function of $B$ over a wide temperature range.\cite{oldtep}
At low temperatures, where $S_{xx}$ for this device is dominated by
phonon drag, no variation of $S_{xx}$ with $B$ up to about 1\,T
was seen to within
the experimental accuracy of $\leq 2\%$ at low fields, even though
changes $\approx 5-6\%$ of $\sigma$ occurred due to localization
effects.

In the present experiments a different cryostat was used with
the sample mounting and associated instrumentation redesigned
to increase the precision of the measurements by about an order of
magnitude in the range 3-7K where phonon drag is dominant in $S_{xx}$,
and also to allow the direction of the magnetic field to be set
either perpendicular or parallel to the 2DEG. The low temperature limit
was chosen to avoid any problems at low magnetic fields
caused by the superconducting/normal transition of In which is used
to provide thermal contact between the sample and cold sink.

The experimental techniques are based on those published
previously.\cite{oldtep} All data were taken by dc methods.
Very slow field sweeps were used for $\rho_{xx}$,
with a sample current for which self-heating was negligible.
To measure the temperature and temperature
gradient we used a matched pair of 22\,k$\Omega$ Philips surface
mount devices which have very high sensitivity at these temperatures
and no observable magnetoresistance (even up to 8T).
During a measurement of $S_{xx}$, $B$ was held constant and the
temperature gradient was repeatedly
switched on and off by routing a fixed power
either to the sample heater, which produced the temperature gradient,
or to a matched heater mounted on the cold sink close to the sample.
In this way the average power input to the sample stage remained constant
and times to equilibrate were short.
With temperature differences mostly in the range of $\sim$100-300mK
we had a
resolution for $\nabla T_x$ of $\le 0.2\%$. It is important to note that
$\nabla T_x$ was independently measured at each value of $B$.
Thermoelectric voltages of 3.5-22$\,\mu$V were
measured with a resolution of a few nV by an EM 11 nanovoltmeter.
The voltages were reproducible to $\le 0.1\%$ when $\nabla T_x$ was
switched on and off at fixed $T$.

During the measurement of $S_{xx}$ the average sample $T$ was
found to increase slightly with $B$. The relative change was
always largest at highest $T$; at 7.4K and 1.5T it reached
a maximum of about 0.8\% for S1, and 0.4\% for S2. On
lowering the field to 0.5T, the changes dropped to 0.2\% for
S1 and 0.1\% for S2. These changes
are similar for both perpendicular and parallel $\bf B$. Each data
point has been corrected to bring them all to the same temperature.
Interestingly, when this is done, the results are essentially the same as
if we had used the zero-field value for $\nabla T_x$ and the as-measured
values for the thermoelectric voltages. This arises because
the thermal conductivity of the substrate and $S_{xx}$ have
similar $T$ dependences. The estimated relative error of
$S_{xx}$ as a function of $B$ is typically $\sim 0.3\%$ and that of
$\rho_{xx}$ is $<0.1\%$. There is a systematic
overall uncertainty in $S_{xx}$ of about 10\% due to possible errors
in thermometer
and voltage contact spacing, but this should not depend on $T$ or $B$.

The temperature dependence of $S_{xx}$ for the two samples
is shown in Fig.~1 along with $S^d_{xx}$ which was measured for
S1\cite{oldtep} but is estimated for S2. The increase in $S_{xx}$ for
S2 compared with S1 is due to the lower carrier density and somewhat
higher substrate thermal conductivity. The data for
S1 are in good agreement with those of previous
measurements.\cite{oldtep}

The upper panels of Figs.~2 and 3 show data on the $B$ dependence of
$\triangle S_{xx}/S_{xx} = (S_{xx}(B)-S_{xx}(0))/S_{xx}(0)$
and $\triangle\sigma_{xx}/\sigma_{xx} =
(\sigma_{xx}(B)-\sigma_{xx}(0))/\sigma_{xx}(0)$
(taking $\sigma_{xx} = 1/\rho_{xx}$)
for the two samples. The plots are given in terms of $\ln
 B$ because
if WL is dominant we expect $\sigma_{xx} \propto \ln B$
over a relatively wide field range.
Although $\triangle\sigma_{xx}/\sigma_{xx}$ is about $6\%$ and $10\%$
respectively for the two samples, the variation of $\triangle
S_{xx}/S_{xx}$ is roughly
an order of magnitude smaller over the same field range,
though it is not zero within experimental error.
These data are contrary to those previously published\cite{Syme1and2}
where $\triangle S_{xx}/S_{xx}$ showed essentially the same
behaviour as $-\triangle \sigma_{xx}/\sigma_{xx}$.

Although there seems no doubt that WL is responsible for
the behaviour of $\sigma_{xx}$, we briefly outline our
analysis of the temperature and field dependence of these data
to show that they do indeed behave as expected.
We expect the Boltzmann value of
$\sigma_{xx}$ (which we write as $\pi\sigma_N k_F l_e$ where
$\sigma_N = e^2/2\pi^2\hbar$, $k_F$ is the magnitude of the
Fermi wave vector and $l_e$ is the elastic mean free path)
to be modified due to contributions from WL,
$\Delta \sigma_{xx}^{WL}$, and
electron-electron interaction, $\Delta \sigma_{xx}^{ee}$,
of the form\cite{wlreviews}
\begin{equation}
\sigma_{xx} = \pi\sigma_N k_F l_e + \Delta \sigma^{WL}_{xx} +
    \Delta \sigma^{ee}_{xx}
   = \sigma_N [\pi k_F l_e - (1 + g) \ln (l_i/l_e)]
\end{equation}
where $\l_i = v_F\tau_i$ is the inelastic mean free path
given in terms of the phase coherence time $\tau_i$,
and $g$ is a measure of the
strength of the electron-electron interaction. The temperature
dependence arises from $l_i$. If one assumes $\l_i \propto 1/T^p$, then
$\sigma_{xx} = p(1+g)\sigma_N  \ln T + $\,constant. Plots of
$\sigma_{xx}$ versus $\ln T$ over the range 2-9\,K for both samples gave
good straight lines with slopes of $1.93\times 10^{-5}\,\Omega^{-1}$
for S1 and $1.85\times 10^{-5}\,\Omega^{-1}$ for S2.
From the analysis of the $B$ dependence given below
we find $p=1.0$; these results imply $g= 0.57$ for S1 and 0.51 for S2,
each $\sim \pm 10\%$.

Noting that $\Delta \sigma^{ee}_{xx}$ is $B$-independent for low
fields, the $B$-dependence of $\sigma_{xx}$ was fitted
to the standard results of WL,\cite{wlreviews} i.e.
\begin{equation}
\label{WLvsB}
 \sigma_{xx} = \sigma_N \left[ \pi k_Fl_e - g \ln \left(\frac{l_i}{l_e}
   \right) + \psi \left(\frac{1}{2} + \frac{\hbar}{2eBl_i l_e}\right) -
    \psi \left( \frac{1}{2} + \frac{\hbar}{2eBl_e^2} \right ) \right]
\end{equation}
where $\psi$ is the digamma function. Using an expression
for the conductivity in this form puts a strong constraint on the
value for $l_e$ because it is required to fit simultaneously both the
Boltzmann term and the correction terms. The usual practice has been
to fit only the field dependent part in isolation.
As shown in Figs.~2 and 3
this expression gives excellent fits to our experimental data in the
range $-0.05$\,T$<B< 0.05$\,T, but at higher fields the fits become
worse and are no longer acceptable for $|B| > 0.1$\,T. This agrees
with expectations since the field dependence in Eq.~(\ref{WLvsB})
is accurate only when $2eBl_e^2/\hbar \ll 1$,
\cite{ZDK97} which corresponds to $|B| \ll 0.25$\,T in our case.
Within experimental error $l_i = a/T$ and $l_e$ is precisely constant
(using fits in the range $-0.05$\,T$< B < 0.05$\,T);
$a$ and $l_e$ are  given in Table~1, along with $k_F l_e$.
If electron-electron effects are ignored in the above ($g$ = 0),
the fitted parameters are changed only weakly.

Our samples are in the dirty limit with $\hbar /\tau_e \gg k_B T$, where
$k_B$ is the Boltzmann constant and $\tau_e$ the impurity relaxation
time. Under these conditions, Altshuler {\it et al.}\cite{AAK82}
have predicted that $l_i$ should be given by
\begin{equation}
\frac{1}{l_i} = \frac{k_B T}{2 \epsilon_F l_e}
      \ln ({\epsilon_F \tau_e}{\hbar})
\end{equation}
where $\epsilon_F$ is the Fermi energy and $\tau_e$ is the
electronic elastic relaxation time. This expression yields
$l_i = 51/T\,\mu$m for S1 and $31/T\,\mu$m for S2, which are to
be compared with the experimental values of $14/T\,\mu$m
and $21/T\,\mu$m. The agreement is certainly not as good as that
obtained by Choi {\it et al.}\cite{Choi87} with GaAs heterojunctions,
but the order of magnitude is correct, especially  for S2. All these
results confirm that our samples are dominated by WL, at least
insofar as the $B$ dependence of $\sigma_{xx}$ is concerned.

In the previous work\cite{Syme1and2} data were also
taken with the magnetic field parallel to the 2DEG, where no change is
expected nor observed for $\sigma_{xx}$, and no change was seen for
$S_{xx}$. We have also done this and typical data on $S_{xx}$
(with ${\bf B} \parallel y$) are shown in the lower panels of
 Figs.~2 and 3 for both samples. Any changes are now
at the level 0.2\% which is within the range of the
expected uncertainties.

\section{Theory}
\label{theory}
In keeping with the experimental situation, we
shall restrict our attention to the case of a 2DEG in which
the electrons occupy a single subband and have an
isotropic energy dispersion. The generalization to multiple
subbands is straightforward but will not be considered in this
paper. Our treatment of the transport properties is based
on the semiclassical Boltzmann equation and is similar to the
approaches used by Butcher's group\cite{Butcher} and Lyo.\cite{Lyo}
It is useful to provide a brief but complete summary of the theory
in order to clearly reveal the underlying approximations that are
made. In addition, our intention is to bring out various features
of the theory, not immediately evident in the previous theoretical
work, which give useful physical insights into the general behaviour
of the phonon drag component of the thermopower in degenerate
semiconductors. A result of particular interest is the
definition of an effective electric field which accounts for the
effect of a nonequilibrium flux of phonons on the electrons.
Once these semiclassical
results have been established, we indicate the way
in which the theory can be extended to include the effects of
weak localization.

In the presence of static electric and magnetic fields, the
electron distribution function $f_{\bf k}({\bf r})$
satisfies the steady-state equation
\begin{equation}
{\bf v}_{\bf k}\cdot \nabla f_{\bf k} - {e\over \hbar}
\left ( {\bf E} + {\bf v}_{\bf k}\times {\bf B} \right ) \cdot
\nabla_{\bf k} f_{\bf k} =
\left.{{\partial f_{\bf k}}\over{\partial t}}
\right\vert_{imp}
+ \left.{{\partial f_{\bf k}}\over{\partial t}}
\right\vert_{ph}
\label{eqn1}
\end{equation}
where the terms on the right hand side account for impurity and
phonon scattering. The impurity scattering term is
given by
\begin{equation}
\left. {{\partial f_{\bf k}}\over{\partial t}}
\right\vert_{imp} = \sum_{\bf k'} W_{\bf kk'} \left [ f_{\bf
k'}(1-f_{\bf k}) - f_{\bf k}(1-f_{\bf k'}) \right ]
\label{eqn2}
\end{equation}
where $W_{\bf kk'}$ is an energy conserving impurity scattering
rate. Similarly, the phonon scattering term is given by\cite{Butcher}
\begin{equation}
\left. {{\partial f_{\bf k}}\over{\partial t}}
\right\vert_{ph} = \sum_{\bf k'}  \left [ f_{\bf
k'}(1-f_{\bf k})P_{\bf k'k} - f_{\bf k}(1-f_{\bf k'}) P_{\bf
kk'}\right ]
\label{eqn3}
\end{equation}
with
\begin{equation}
P_{\bf kk'} = {2\pi \over \hbar} \sum_Q |M_{\bf kk'}(Q)|^2 \left
[ N_Q \delta(\varepsilon_{\bf k'} - \varepsilon_{\bf k} -\hbar
\omega_Q)+(N_{-Q}+1) \delta(\varepsilon_{\bf k'} -
\varepsilon_{\bf k} +\hbar\omega_Q) \right ]\,.
\label{eqn4}
\end{equation}
The variable $Q$ represents both the three-dimensional phonon
wavevector ${\bf Q}$ with components parallel (${\bf Q}_\parallel$)
and perpendicular ($Q_\perp$) to the 2DEG,
and polarization index $\lambda$; $\omega_Q$ is the phonon
frequency and $N_Q$ is the nonequilibrium phonon occupation
number. The term proportional to $N_Q$ corresponds to phonon
absorption whereas the $(N_{-Q}+1)$ term corresponds to
emission. The electron-phonon matrix element has the property
$M_{\bf kk'}(Q) = M_{\bf k'k}^*(-Q)$ and is proportional
to $\delta_{{\bf k'},{\bf Q}_\parallel + {\bf k}}$, due to
momentum conservation in the plane of the 2DEG, and to the
subband matrix element $I(Q_\perp) \equiv \int
\exp(iQ_\perp z)|\phi(z)|^2 dz$.
The detailed form of $M_{\bf kk'}(Q)$ depends on the nature of
the interaction; the acoustic phonon deformation potential and
piezoelectric interactions are relevant to the GaAs/GaAlAs systems
studied here. Finally it should be noted that electronic
screening of the electron-phonon interaction is
included by dividing the bare electron-phonon matrix element by
the dielectric function $\epsilon({\bf Q}_\parallel)$ of the 2DEG.

The solution to Eq.~(\ref{eqn1}) is obtained by linearizing about
a local equilibrium ($le$) Fermi distribution, i.e.
\begin{equation}
f_{\bf k}({\bf r})= f_{\bf k}^{le}(T({\bf r}),\mu({\bf r})) -
{\partial f^0_{\bf k} \over \partial \varepsilon_{\bf k}}
\Phi_{\bf k}\,,
\label{eqn5}
\end{equation}
where $T({\bf r})= T+\delta T({\bf r})$ is the local temperature
and $\mu({\bf r}) = \mu +\delta \mu({\bf r})$ is the
local chemical potential. The thermodynamic equilibrium distribution
at temperature $T$ and chemical potential $\mu$ will be denoted by
$f^0_{\bf k}=(e^{\beta(\varepsilon_{\bf k} -\mu)}+1)^{-1}$.
The deviation from local equilibrium is written
in the form shown for later convenience.
Similarly, the phonon distribution is expanded as
\begin{equation}
N_Q= N_Q^{le} - {\partial N_Q^0 \over \partial (\hbar \omega_q)}
G_Q\,,
\label{eqn6}
\end{equation}
where $N_Q^0=(e^{\beta\hbar\omega_Q}-1)^{-1}$
is the equilibrium Bose distribution at temperature
$T$. We shall assume that the nonequilibrium phonon distribution
$G_Q$ is established by a temperature gradient and is
insensitive to the interactions of the phonons with the electronic
system. However,  phonon-drag thermopower arises directly
from the interaction of the electrons with the nonequilibrium
distribution of phonons.

Substituting these forms of the electron and phonon distributions
into Eq.~(\ref{eqn1}), we obtain the linearized equation
(details of the derivation of the phonon scattering terms
can be found in the work of Cantrell and Butcher\cite{Butcher})
\begin{equation}
{\bf v}_{\bf k}\cdot \nabla T \,{\partial f_{\bf k}^0 \over
\partial T} -
e{\bf v}_{\bf k}\cdot {\bf E'} \,{\partial f^0_{\bf k} \over \partial
\varepsilon_{\bf k}} + {e\over \hbar} {\bf v}_{\bf k}\times
{\bf B} \cdot \nabla_{\bf k} \Phi_{\bf k} \,{\partial f^0_{\bf k}
\over \partial\varepsilon_{\bf k}}
=
\left. {{\partial f_{\bf k}}\over{\partial t}}
\right\vert_{imp}
+\left. {{\partial f_{\bf k}}\over{\partial t}}
\right\vert_{ph}
\label{eqn7}
\end{equation}
{\noindent
where
}
\begin{equation}
\left. {{\partial f_{\bf k}}\over{\partial t}}
\right\vert_{imp} = -\sum_{\bf k'} W_{{\bf kk'}}\left [\Phi_{\bf
k'} - \Phi_{\bf k}\right ] {\partial f^0_{\bf k} \over \partial
\varepsilon_{\bf k}}
\label{eqn8}
\end{equation}
and
\begin{equation}
\left. {{\partial f_{\bf k}}\over{\partial t}} \right\vert_{ph} =
\beta \sum_{{\bf k'},Q} \left \{ \Phi_{\bf k'} - \Phi_{\bf k}\right
\} \left [\Gamma_{{\bf kk'}}(Q)+\Gamma_{{\bf
k'k}}( Q)\right ]
-\beta\sum_{{\bf k'},Q} G(Q)
\left [\Gamma_{{\bf kk'}}(Q)-\Gamma_{{\bf k'k}}(Q)
\right ]
\label{eqn9}
\end{equation}
{\noindent
with}
\begin{equation}
\Gamma_{{\bf kk'}}(Q) = {2\pi \over \hbar} |M_{\bf
kk'}(Q)|^2 f^0_{\bf k} (1-f^0_{\bf k'}) N^0_Q
\delta(\varepsilon_{\bf k'} - \varepsilon_{\bf k} - \hbar
\omega_Q)\,.
\label{eqn10}
\end{equation}
The effective electric field ${\bf E'}$ in Eq.~(\ref{eqn7})
is now the gradient
of the electrochemical potential, ${\bf E}+\nabla \mu/e$.
The first sum on the right hand side of Eq.~(\ref{eqn9}), which
we call
$\partial f_{\bf k}/\partial t\vert_{ph,1}$, represents the
scattering of the nonequilibrium distribution of electrons from
the thermal distribution of phonons which has the effect of
equilibrating the electrons to the lattice. The second sum,
$\partial f_{\bf k}/\partial t\vert_{ph,2}$,
accounts for the effect of the nonequilibrium phonons in driving
the electrons out of equilibrium.
When considered in isolation, the first sum is responsible
for the resistivity due to phonon scattering.
However, for the samples of interest here, impurity scattering
is the dominant scattering mechanism at low temperatures
and this term can be neglected in comparison to Eq.~(\ref{eqn8}).
The equilibrium-phonon scattering term is nevertheless of
interest as will be discussed later.

We now concentrate on the nonequilibrium phonon scattering term,
$\partial f_{\bf k}/\partial t\vert_{ph,2}$, which is
responsible for phonon-drag thermopower. It depends on the
nonequilibrium phonon distribution which is obtained from the
phonon kinetic equation
\begin{equation}
{\bf v}_{ph}(Q)\cdot \nabla N_Q = \left. {{\partial N_Q
}\over{\partial t}} \right\vert_{coll}\equiv -{1\over
\tau_{ph}}\left (N_Q -N_Q^{le} \right )\,,
\label{eqn11}
\end{equation}
where the phonon relaxation time $\tau_{ph}$ is a parameter
characterizing boundary and impurity scattering. In principle
it is a function of $Q$ but we will take it to be a simple
$Q$-independent constant which can be obtained from the phonon
thermal conductivity. With our earlier definition of $G_Q$,
Eq.~(\ref{eqn11}) implies that
\begin{equation}
G_Q  = - {\hbar \omega_Q \over T} \tau_{ph} {\bf v}_{ph}(Q)\cdot
\nabla T\,.
\label{eqn12}
\end{equation}
Within an isotropic Debye model, the phonon velocity
${\bf v}_{ph}(Q) = s_\lambda \hat {\bf Q}$ has distinct values
for the longitudinal and transverse modes.

The distribution function in Eq.~(\ref{eqn12}) has the property $G(Q) =
-G(-Q)$, which is the asymmetry one would expect
to see in the presence of
a temperature gradient. Using this property, the nonequilibrium
phonon scattering term in Eq.~(\ref{eqn9}) can be written in the form
\begin{equation}
\left. {{\partial f_{\bf k}}\over{\partial t}} \right\vert_{ph,2} =
- \beta\sum_{{\bf k'},Q} G(Q)
\left [\Gamma_{{\bf kk'}}(Q)+\Gamma_{{\bf k'k}}(-Q)
\right ]\,,
\label{eqn13}
\end{equation}
and substituting Eq.~(\ref{eqn12}) into Eq.~(\ref{eqn13}), we find
\begin{equation}
\left. {{\partial f_{\bf k}}\over{\partial t}} \right\vert_{ph,2} =
{1\over k_B T^2} \sum_{{\bf k'},Q} (\hbar\tau_{ph}
s_\lambda^2)\nabla T \cdot {\bf Q}_\parallel
\left [\Gamma_{{\bf kk'}}(Q)+\Gamma_{{\bf k'k}}(-Q)
\right ]\,.
\label{eqn14}
\end{equation}
The appearance of the in-plane projection ${\bf Q}_\parallel$ in
the sum assumes that the temperature gradient is in the plane of
the 2DEG.

To simplify Eq.(\ref{eqn14}) it is convenient to make use of the
following identities
\begin{eqnarray}
f^0(\varepsilon)[1-f^0(\varepsilon+\hbar \omega)]N^0(\omega)
&=& -{1\over \beta} {\partial f^0 \over \partial
\varepsilon} \left[ N^0(\omega)+f^0(\varepsilon+\hbar
\omega) \right ] \nonumber \\
f^0(\varepsilon-\hbar \omega)[1-f^0(\varepsilon)]N^0(\omega)
&=& -{1\over \beta} {\partial f^0 \over \partial
\varepsilon} \left[ N^0(\omega)+1-f^0(\varepsilon-\hbar
\omega) \right ]\,,
\label{eqn16}
\end{eqnarray}
for the thermal factors appearing in the scattering rates
$\Gamma_{{\bf kk'}}(Q)$ defined in Eq.~(\ref{eqn10}).
With these results, together with
the momentum conservation condition ${\bf Q}_\parallel =
{\bf k'} - {\bf k}$, Eq.(19) can be written in the form
\begin{equation}
\left. {{\partial f_{\bf k}}\over{\partial t}} \right\vert_{ph,2} =
{\partial f^0_{\bf k} \over \partial \varepsilon_{\bf k}}
\sum_\lambda {m^* s_\lambda \Lambda_\lambda \over
\tau_{ep}^\lambda(\varepsilon)} {\bf v}_{\bf k} \cdot {\nabla T
\over T}\,,
\label{eqn19}
\end{equation}
where $\Lambda_\lambda = s_\lambda \tau_{ph}$ is the phonon mean
free path and $\tau_{ep}^\lambda(\varepsilon)$ is an
energy-dependent electron-phonon relaxation time defined as
\begin{eqnarray}
{1 \over \tau_{ep}^\lambda(\varepsilon) }  &=&
{2\pi \over \hbar} \sum_{\bf k',Q} \left
( 1 - {k'\over k}\cos \theta \right ) |M_{\bf kk'}(Q)|^2 \nonumber \\
&\times& \left \{ \left [ N_Q^0 + f^0(\varepsilon')\right ]
\delta(\varepsilon' - \varepsilon - \hbar \omega_Q) + \left [
[ N_Q^0 +1 - f^0(\varepsilon')\right ] \delta(\varepsilon' -
\varepsilon + \hbar \omega_Q) \right \}
\label{eqn17b}\,.
\end{eqnarray}
Here, $\theta$ is the scattering angle between the initial
wavevector ${\bf k}$ and the final wavevector ${\bf k}'$ and
we have introduced the short-hand notation $\varepsilon
\equiv \varepsilon_{\bf k}$ and $\varepsilon' \equiv
\varepsilon_{\bf k'}$.
Provided that Eq.(\ref{eqn12}) is an accurate representation of
the nonequilibrum phonon distribution, Eq.(\ref{eqn19}) is a
rigorous result for the phonon collision integral.

If we now define the effective electric field
\begin{eqnarray}
{\bf E}_{ph}(\varepsilon) = \sum_\lambda {m^* s_\lambda
\Lambda_\lambda \over
e\tau_{ep}^\lambda(\varepsilon)} {\nabla T \over T}\,,
\label{eqn20}
\end{eqnarray}
Eq.~(\ref{eqn19}) can be written in the suggestive form
\begin{equation}
\left. {{\partial f_{\bf k}}\over{\partial t}} \right\vert_{ph,2}
= e {\bf v}_{\bf k} \cdot {\bf E}_{ph}(\varepsilon) {\partial
f^0_{\bf k} \over \partial \varepsilon_{\bf k}}\,.
\label{eqn20a}
\end{equation}
In the context of the original Boltzmann equation in
Eq.~(\ref{eqn7}), the electron-phonon
scattering term given by Eq.~(\ref{eqn20a}) (as stated
earlier, we neglect the
$\partial f_{\bf k} /\partial t\vert_{ph,1}$ term)
can be grouped together
with the term in Eq.~(\ref{eqn7}) arising from the actual electric
field ${\bf E}'$. In other words, the phonon-drag field, ${\bf
E}_{ph}(\varepsilon)$, proportional to the temperature gradient,
has exactly the same effect on the nonequilibrium electron
distribution as the actual electric field. Its dependence on
energy through the electron-phonon relaxation time
is only an apparent complication. Since the dominant impurity
scattering mechanism is elastic, the impurity Boltzmann
equation can be solved at a given energy $\varepsilon$ to define
an energy-dependent impurity conductivity
$\mathop{\sigma}\limits^{\leftrightarrow}(\varepsilon;B)$.
The contribution to the
current density arising from the phonon-drag field can then be
expressed in the form
\begin{equation}
{\bf J}_{ph} = \int d\varepsilon \left( - {\partial f^0 \over
\partial \varepsilon} \right )
\mathop{\sigma}\limits^{\leftrightarrow}(\varepsilon;B)
\,{\bf E}_{ph}(\varepsilon)\,.
\label{eqn21}
\end{equation}
It can be shown that this result for the current density is
equivalent to that given by Zianni {\it et al.}\cite{Zianni} and
therefore leads to the same results for the thermoelectric
tensor and phonon-drag thermopower.

Our interest here is in the low temperature properties of a
degenerate 2DEG.
Since the impurity conductivity at low magnetic fields depends
only weakly on energy, it can be taken out of the integral
in Eq.(\ref{eqn21}) and we
simply obtain the energy average of the effective field,
\begin{equation}
{\bf J}_{ph} =
\mathop{\sigma}\limits^{\leftrightarrow}(B)
\,\langle {\bf E}_{ph} \rangle\,,
\label{eqn21a}
\end{equation}
where $\mathop{\sigma}\limits^{\leftrightarrow}(B)$ is the
measurable conductivity appearing in Eq.~(\ref{Ohm}). In view
of Eqs.~(\ref{eqn20}) and (\ref{eqn21a}), the thermoelectric
tensor is
\begin{equation}
\mathop{\epsilon^g}\limits^{\leftrightarrow}(B) =
- \sum_\lambda {m^* s_\lambda \Lambda_\lambda \over
e T} \left \langle {1\over \tau_{ep}^\lambda} \right
\rangle \mathop{\sigma}\limits^{\leftrightarrow}(B)\,,
\label{eqn22}
\end{equation}
where the average lifetime is given by
\begin{eqnarray}
\label{eqn28}
\left \langle {1\over \tau_{ep}^\lambda} \right \rangle \equiv
\int d\varepsilon
\left (- {\partial f^0 \over \partial \varepsilon}
\right ) {1\over \tau_{ep}^\lambda(\varepsilon)}\,.
\end{eqnarray}
The phonon drag thermopower tensor, defined as
$\mathop{S^g}\limits^{\leftrightarrow}(B) =
{\mathop{\sigma}\limits^{\leftrightarrow}}^{-1}(B)
\mathop{\epsilon^g}\limits^{\leftrightarrow}(B)$, is therefore
given by $\mathop{S^g}\limits^{\leftrightarrow}(B) =
S^g \mathop{1}\limits^{\leftrightarrow}$ with
\begin{equation}
S^g = - \sum_\lambda {m^* s_\lambda \Lambda_\lambda \over
eT} \left \langle {1\over \tau_{ep}^\lambda} \right \rangle \,.
\label{eqn23}
\end{equation}
We see that the phonon drag thermopower is a diagonal tensor
{\it independent} of the magnetic field. This result is a direct
consequence of the thermoelectric tensor being a scalar multiple
of the conductivity.  Equations (\ref{eqn20}) through
(\ref{eqn23}) are the main results of the theory.

An equation similar to Eq.~(\ref{eqn23}) was first given by
Herring\cite{Herring} for 3D non-degenerate semiconductors,
and is explained in terms of the following physical picture.
The non-equilibrium phonons preferentially flow down the
temperature gradient $\nabla T$ carrying a crystal momentum
current $\propto \Lambda\nabla T$. Scattering of electrons
by phonons leads to momentum being transferred to the electrons
at a rate proportional to $1/\tau_{ep}$, where
$\tau_{ep}$ is the {\it e-p} momentum relaxation time. This
acceleration of the electrons proceeds for a mean time determined by
the {\it e-i} momentum relaxation time $\tau_{ei}$, (we are
assuming, as above, that $\tau_{ep} \ll \tau_{ei}$) at which
point impurity scattering randomizes the momentum. The mean drift
velocity of the electrons established in this way gives rise to the
electric current contribution ${\bf J}_{ph}$.
Thus ${\bf J}_{ph}  = -\epsilon^g \nabla T \propto
(\tau_{ei}/\tau_{ep})\Lambda \nabla T$.
Since the thermopower is measured with ${\bf J} =0$,
a compensating drift current, ${\bf J}_\sigma = \sigma {\bf E}$,
is established. With ${\bf J}_{ph} + {\bf J}_\sigma
 = 0$ and $\sigma  \propto \tau_{ei}$, the induced electric
field is ${\bf E} \propto  \Lambda\nabla T/ \tau_{ep}$, and
$S^g$ has the form given in Eq.~(\ref{eqn23}).

Before proceeding, we would like to indicate the usefulness of
the results we have obtained so far. In the Appendix we show
that the electron-phonon relaxation time appearing in
Eq.~(\ref{eqn23}) is closely related to the electron-phonon
transport lifetime that arises in the context of the phonon-limited
mobility. The latter is defined by
$\mu_{ep} \equiv e\langle 1/\tau_{tr} \rangle^{-1}/m^*$ where
the transport lifetime is given in Eq.~(\ref{eqn17c}) of the Appendix.
At low temperatures, $\langle 1/\tau_{tr}^\lambda \rangle$ and $\langle
1/\tau_{ep}^\lambda \rangle$ are essentially the same, in which
case the phonon-drag thermopower can be expressed as $S^g \simeq
-\sum_\lambda (s_\lambda \Lambda_\lambda/
\mu_{ep}^\lambda T)$ with $\mu_{ep}^\lambda =
e\langle 1/\tau_{tr}^\lambda \rangle^{-1}/m^*$. This result
has been applied to degenerate 2DEGs (e.g. see Ref.
\onlinecite{approx}),
but it has been viewed as a semi-quantitative result, which indeed
is the case for non-degenerate semiconductors. The fact that it is
quantitatively accurate for the degenerate case (at least in the low
temperature limit)
was discovered empirically in recent experimental work\cite{Tieke}
concerned with the thermopower of 2DEGs and composite fermions,
and was used there to evaluate
$\mu_{ep}$ for these systems. The importance of this connection
is that the contribution to the resistivity from phonon scattering
can be very small and difficult to measure at low temperatures,
even in high mobility systems. Through Eq.~(\ref{eqn23}), the
thermopower provides an alternate way to measure
$\mu_{ep}$ which can be much more accurate.

Equation~(\ref{eqn23}) shows that the longitudinal phonon
drag thermopower $S^g_{xx}$ should be independent of magnetic
field, and that the phonon drag contribution to the
Nernst-Ettingshausen coefficient $S^g_{yx}$ should be zero.
These results are contained in the theoretical work of Zianni
{\it et al.}\cite{Zianni} but were presented in a form which
is not as physically transparent as that given here. The
field-independence of $S^g_{xx}$ has been demonstrated
experimentally in a number of 2DEG systems, e.g.
Refs. \onlinecite{oldtep} and \onlinecite{MOSFET}, though it is also
evident in much of the earlier work. However in contrast to the
theoretical prediction, it was found that $S^g_{yx}$ is not zero
for the same systems. Butcher and Tsaousidou\cite{Rita}
have suggested anisotropy of the electrons and phonons
as a possible reason for this discrepancy, but this explanation
would also imply some field variation of $S^g_{xx}$.

Although the above results were derived with 2DEGs in mind,
it should be emphasized that they are equally valid in
three dimensions when the same physical conditions prevail. In
Eq.~(\ref{eqn14}) for example, the summation over ${\bf k}'$
is simply interpreted as a three-dimensional sum, all
electron-phonon matrix elements are evaluated using
three-dimensional plane wave states and ${\bf
Q}_\parallel$ is replaced by the total phonon wave vector
${\bf Q}$. The subsequent analysis leading to the thermopower in
Eq.~(\ref{eqn23}) remains unchanged. It was recently demonstrated
experimentally that $S^g_{yx} = 0$ to high accuracy in a degenerate
3D semiconductor, even when $S^g_{xx}$ is completely dominant
in the longitudinal case,\cite{HGSE} and that  $S^g_{xx}$ is
independent of magnetic field. In this respect, the 3D situation
conforms even more closely to the theoretical predictions.

We now turn to the aspect of immediate concern in this paper,
that of WL.  The weak field dependence shown by
our experimental results for the thermopower
is consistent with the conclusions reached above on
the basis of the semiclassical Boltzmann equation, even though
the behaviour of the conductivity in our samples is manifestly
nonclassical. We argue that these observations have a
natural explanation in terms of a generalized Boltzmann
equation developed by Hershfield and Ambegaokar.\cite{Hersh}
These authors showed that the effects of coherent backscattering
can be included with the addition of an extra term in
the semiclassical Boltzmann equation. Although their final
results are derived in the absence of a magnetic field, they
claim that their approach can be extended to include a
magnetic field and that all the standard WL results for
the magnetoresistance can be reproduced. The significance of
this is that WL effects are in principle
accessible within an otherwise semiclassical approach.

We accept the argument that a suitable modification of the
impurity scattering term in Eq.~(\ref{eqn1})
can be made which, in the absence of phonon scattering, leads to
a conductivity with WL effects included, and now
consider the additional effects of phonon scattering.
This is one inelastic scattering mechanism contributing to the
phase-coherence lifetime $\tau_i$ appearing in the WL
correction, though electron-electron scattering is usually
the dominant mechanism at low temperatures. Of interest here
are the additional effects of phonon collisions which impart
momentum to the electrons and induce a nonequilibrium electron
distribution.

Syme {\it et al.}\cite{Syme3,Syme4} argued that, since phonon
scattering events are phase disrupting, and since phonon drag
originates in these events, $\epsilon^g_{xx}$
will not exhibit WL effects.  We believe that
this is incorrect. Within a semiclassical description of the
electron dynamics, we have shown that the effect of
phonon collisions is equivalent to an effective electric field
in so far as the subsequent current response is concerned.
Whether the impulse is provided by an electric field
or a phonon collision, the induced current is limited by
impurity scattering and is proportional to the impurity conductivity
in either case. The same conclusion should also apply in the
WL regime when $\tau_{ep} \gg \tau_{ei}$.
Once an impulse has been applied,
an electron propagates in the presence of the impurities and the
full effects of quantum interference are operative up to a time
determined by $\tau_i$.
It is important to emphasize that the frequency
of electron-phonon scattering events is unchanged when the phonon
distribution is displaced from equilibrium by the temperature
gradient. This means that the frequency of phase disrupting
events is unchanged and the relevant conductivity determining
the current response is the one including the WL effects. Thus
with the use of the generalized Boltzmann equation to incorporate
WL effects, our final results for the phonon drag thermoelectric
tensor given in Eq.~(\ref{eqn22}) should
still be valid with the simple replacement of the conductivity
by the one including weak-localization corrections.

As a final comment, we note that the above argument used for
phonon drag
can also be made in the case of the diffusion thermopower.
The first term on the left hand side of Eq.~(\ref{eqn7})
gives rise to the diffusion current
\begin{equation}
{\bf J}_{d} = \frac{1}{e}
\int d\varepsilon \left( - {\partial f^0 \over
\partial \varepsilon} \right ) (\varepsilon-\mu)
\mathop{\sigma}\limits^{\leftrightarrow}(\varepsilon;B)
\,{\nabla T \over T} \,.
\label{eqn21b}
\end{equation}
Using a conductivity with WL corrections in this expression
leads to the accepted form of the diffusion thermopower which, in
contrast to the phonon-drag thermopower, does exhibit WL
corrections.\cite{Afonin84,Castellani88,KandB88,Hsu89}

To summarize, we have argued that the role of phonon drag in
establishing an electric current is equivalent to an applied
electric field in the impurity-dominated regime, with the
consequence that the phonon drag thermoelectric tensor is
proportional to the impurity conductivity. The WL corrections
within the conductivity also appear in the
thermoelectric tensor, resulting in a cancellation of WL effects
in phonon drag thermopower. These results are essentially in
accord with the present experimental data.

\section{Discussion}
The origin of the discrepancies between the present and
previous work with regard to $S^g_{xx}$ in a perpendicular
magnetic field is not clear. The fact that the previous data
were obtained on a MOSFET and the present data are for
GaAs/Ga$_{1-x}$Al$_x$As quantum wells should be irrelevant
if WL is the cause of the field dependence of $S^g_{xx}$.
This in itself is significant
because if the differences are real, it implies that something other
than weak localization is involved in at least one of these experiments.
One possibility might be that the previous samples had
$k_F l_e \sim 2-4$,
though WL theory is expected to be valid only for $k_F l_e \gg 1$.
The present samples have $k_F l_e > 10$ (Table~1). Much less is
known, both experimentally and theoretically, about the situation
at small $k_F l_e$ where the system is approaching strong localization.
Ashe {\it et al.}\cite{Ashe92} have probed
$\sigma_{xx}$ in this limit but were unable to find satisfactory
agreement with WL, even with extensions to the
theory. How phonon drag might behave in this limit is unknown.

There are also some features of the experimental technique
used in the previous work that might be cause for concern.
 Syme {\it et al.}\cite{Syme1and2,Syme3,Syme4} used an ac
method for which the temperature difference could be measured
only when $B=0$ and it was assumed to remain unchanged at finite $B$.
Also, although constantan potential leads were used to
minimize heat losses when thermal conductivity was
measured, these were replaced by Cu leads for $S_{xx}$.
The authors argued that this gives isothermal conditions
in the direction transverse to the applied temperature
gradient. However, the use of Cu leads implies a significant
heat leak to the surroundings and unknown thermal
boundary conditions on the 2DEG; this might
be particularly important with an AC measurement where
the heat flow is continuously changing.

Although the change we see in $S^g_{xx}$ is much smaller than that in
$\sigma_{xx}$, nevertheless it is outside experimental error,
especially for S2, and we must ask how it arises.
An obvious possibility is that it is
due to diffusion. The effect of WL on $S^d_{xx}$,
say $\Delta S^d_{xx}$, is believed to be
\cite{Afonin84,Castellani88,KandB88,Hsu89}
\begin{equation}
  \frac{\Delta S^d_{xx}}{S^d_{xx}} \approx
                        \frac{\ln (l_i/l_e) }{\pi k_F l_e}\, .
\end{equation}
For our samples at 3\,K (7\,K),
$\ln (l_i/l_e) /\pi k_F l_e = 0.08$ (0.06) for S1, and 0.12 (0.01)
for S2. From Fig.~1 we see that for S1 (S2) the measurements show
that $S^g_{xx}/S^d_{xx}$ varies between about 25-70 (40-65)
for the temperature range 3-7K. This suggests that
changes due to this cause would be $\sim 0.3\%$ at 3K
and $\sim 0.1\%$ at 7K. These estimates agree with
the observations for S1, where $\Delta S_{xx}/S_{xx}$ is
very small, but seem rather too low to account
for the variations for S2. We would expect the effect to decrease
as temperature increases thereby decreasing the relative
importance of $S^d_{xx}$, but the experimental results
are not unambiguous on this point.
We also note that these effects might be much larger,
and basically unknown, for the samples of
Syme {\em et al.}\cite{Syme1and2,Syme3} with $k_F l_e \sim 2-4$.

Another possibility is that we may be observing the effects of
anisotropy in the electron and phonon system. As we noted
in Section \ref{theory}, Butcher and Tsaousidou\cite{Rita} have
suggested this as the origin of the finite $S^g_{yx}$ observed
in 2DEGs. If this is correct, it will be accompanied by a
field dependence of $S^g_{xx}$, and this should be essentially
independent of temperature, not inconsistent with the
observations.

\section{Conclusions}

Our measurements show that the longitudinal phonon drag thermopower
$S^g_{xx}$ is essentially independent of magnetic field, which
implies that it is independent of weak localization, a result
contrary to earlier work. We have given
a theory of phonon drag based on the semiclassical Boltzmann approach
which shows that $S^g_{xx}$ is independent of magnetic field. Using
the results of Hershfield and Ambegaokar\cite{Hersh},
we have argued that this result
remains valid in the presence of weak localization. Our theory also
gives results which are useful in understanding the origin and
behaviour of phonon drag thermopower in 2D and 3D degenerate
systems.

\acknowledgments

The work was supported by grants from the Natural Sciences and
Engineering Research Council of Canada.

\appendix
\section{}
It is apparent that Eq.~(\ref{eqn14}) bears a strong similarity
to the first
phonon scattering term in Eq.~(\ref{eqn9}) which is dependent on
the nonequilibrium electron distribution. This similarity can be
exploited to provide a relation between the phonon drag driving
term given by Eq.~(\ref{eqn14}) and the electron-phonon momentum
relaxation time which determines the phonon-limited mobility.
This latter quantity is obtained by solving the following transport
equation
\begin{equation}
e{\bf v}_{\bf k}\cdot {\bf E} {\partial f^0_{\bf k} \over \partial
\varepsilon_{\bf k}}
=
- \beta \sum_{{\bf k'},Q} \left \{ \Phi_{\bf k'} - \Phi_{\bf k}\right
\} \left [\Gamma_{{\bf kk'}}(Q)+\Gamma_{{\bf
k'k}}(-Q)\right ]\,,
\label{eqn15}
\end{equation}
which describes the transport of electrons scattering exclusively from
phonons, that is, in the absence of impurities. It should also
be noted that we have not included a magnetic field or thermal
gradient in this equation. This of course
is not the physical situation of interest in Eq.~(\ref{eqn7}),
but rather represents an auxiliary problem which identifies a useful
transport property that is relevant to the phonon-drag
calculation. In addition, a discussion of Eq.~(\ref{eqn15}) will
allow us to make contact with earlier
work\cite{Price,Stormer,kawamura}
on the calculation of electron-phonon mobilities.

The right hand side of Eq.~(\ref{eqn15}) follows from Eq.~(\ref{eqn9})
with the replacement $Q \to -Q$ in the $\Gamma_{{\bf k'k}}(Q)$
term. This puts it in the same form as the corresponding term in
Eq.~(\ref{eqn14}). To proceed with the solution of
(\ref{eqn15}), we make use of the identities in
Eq.(\ref{eqn16}). The appearance of the
Fermi function derivatives allows the transport equation to be
simplified as
\begin{eqnarray}
e{\bf v}_{\bf k}\cdot {\bf E} =
{2\pi \over \hbar} \sum_{{\bf k'},Q}
\left \{ \Phi_{{\bf k'}} - \Phi_{\bf k}\right \}
|M_{\bf kk'}(Q)|^2
&\Bigl \{& \left [ N_Q^0 + f^0(\varepsilon')
\right ] \delta(\varepsilon' -
\varepsilon - \hbar \omega_Q) \nonumber \\
&+& \left [
[ N_Q^0 +1 - f^0(\varepsilon')\right ]
\delta(\varepsilon' -
\varepsilon + \hbar \omega_Q) \Bigr \}\,.
\label{eqn17}
\end{eqnarray}
We now look for a solution to this equation which has the form
$\Phi_{\bf k}
= - e{\bf v}_{\bf k}\cdot {\bf E} \tau_{tr}(\varepsilon)$.
We shall refer to
$\tau_{tr}(\varepsilon)$ as the electron-phonon transport
lifetime to distinguish it from the electron-phonon relaxation
time defined in Eq.~(\ref{eqn17b}). With this
ansatz, Eq.~(\ref{eqn17}) becomes
\begin{eqnarray}
1 =
{2\pi \over \hbar} \sum_{{\bf k'},Q}
\Bigl \{ \tau_{tr}(\varepsilon) - {k' \cos \theta \over
k} \tau_{tr}(\varepsilon') \Bigr \}
|M_{\bf kk'}(Q)|^2
&\Bigl \{& \left [ N_Q^0 + f^0(\varepsilon'
)\right ] \delta(\varepsilon' -
\varepsilon - \hbar \omega_Q) \nonumber \\
&+& \left [
[ N_Q^0 +1 - f^0(\varepsilon')\right ]
\delta(\varepsilon' -
\varepsilon + \hbar \omega_Q) \Bigr \}
\label{eqn17a}\,.
\end{eqnarray}
This is an integral equation for $\tau_{tr}(\varepsilon)$
and is equivalent to Eq.~(22) in
Ref.~[~\onlinecite{kawamura}~]. The form of Eq.~(\ref{eqn17a})
is obviously very similar to Eq.~(\ref{eqn17b}).
It simplifies if we now make the quasielastic
approximation\cite{kawamura} which assumes
that the energies of interest are near the
Fermi energy $\varepsilon_F$ and $\hbar \omega_Q \ll
\varepsilon_F$. We can then replace $\varepsilon'$ by $\varepsilon$
in the $\tau$-dependent factor in Eq.~(\ref{eqn17a}), and
we obtain the following explicit expression for the
relaxation time:
\begin{eqnarray}
{1 \over \tau_{tr}(\varepsilon) }  &=&
{2\pi \over \hbar} \sum_{{\bf k'},Q}
( 1 - \cos \theta ) |M_{\bf kk'}(Q)|^2 \nonumber \\
&\times& \left \{ \left [ N_Q^0 + f^0(\varepsilon')\right ]
\delta(\varepsilon' - \varepsilon - \hbar \omega_Q) + \left [
[ N_Q^0 +1 - f^0(\varepsilon')\right ] \delta(\varepsilon' -
\varepsilon + \hbar \omega_Q) \right \}
\label{eqn17c}\,.
\end{eqnarray}
It can be shown that this expression for $\tau_{tr}(\varepsilon)$
is identical to Eq.~(39) in Ref.~[\onlinecite{kawamura}]
where the quasielastic approximation is also invoked.
The summation over $Q$ includes the
polarization index $\lambda$ so that $1/\tau_{tr}(\varepsilon) =
\sum_\lambda 1/\tau_{tr}^\lambda(\varepsilon)$. It is clear that
if the quasielastic approximation is also made in Eq.~(\ref{eqn17b}),
$1/\tau_{ep}^\lambda(\varepsilon)$  and
$1/\tau_{tr}^\lambda(\varepsilon)$ would in fact be
identical. We expect any quantitative differences to be minimal
at low tempertures where the two relaxation times can be used
interchangeably.

Finally, we can make contact with the average electron-phonon
scattering rate which was derived by Stormer {\it et al.}\cite{Stormer}
on the basis of the work of Price\cite{Price}.
If Eq.~(\ref{eqn17c}) is averaged over energy with respect
to the weight function $(-\partial f^0 / \partial \varepsilon)$,
we obtain the result
\begin{eqnarray}
\left \langle {1\over \tau_{tr}} \right \rangle
= {2\pi \over \hbar} \sum_{{\bf k'},Q} (1&-&\cos \theta)
|M_{\bf kk'}(Q)|^2 \beta \{ f^0(\varepsilon')
[1-f^0(\varepsilon' +\hbar \omega_Q)] N_Q^0 \nonumber \\
&+& f^0(\varepsilon')
[1-f^0(\varepsilon' -\hbar \omega_Q)] (N_Q^0 + 1) \}\,.
\label{eqn18a}
\end{eqnarray}
In arriving at this expression, we have again used the
identities in Eq.~(\ref{eqn16}) as well as the identity
$f^0(\varepsilon - \hbar \omega)[1-f^0(\varepsilon)] N^0(\omega) =
f^0(\varepsilon)[1-f^0(\varepsilon - \hbar
\omega)](N^0(\omega)+1)$. The result actually given by Stormer
{\it et al.}\cite{Stormer} follows from Eq.(\ref{eqn18a}) by
performing the ${\bf Q}_\parallel$ sum with the understanding
that $k = k'$, so that $Q_\parallel = 2k'\sin(\theta/2)$.
The phonon-limited mobility defined as $\mu_{ep} = e\langle
1/\tau_{tr} \rangle^{-1}/m^*$ was used by Stormer
{\it et al.}\cite{Stormer} to interpret their observation of a
transition into the Bloch-Gr\"uneisen regime of the
mobility of a 2DEG. Eq.~(\ref{eqn18a}) is applicable at
arbitrary temperatures, but a simpler expression can be obtained
in the limit of low temperatures.  Since the derivative of the
Fermi function is then sharply peaked at the
chemical potential $\mu$, the average in
Eq.~(\ref{eqn18a}) may be replaced by the value at the chemical
potential if $\tau_{tr}(\varepsilon)$ is a slowly varying
function of energy. In this case we have
(see Eq.~(\ref{eqn17c}))
\begin{equation}
{1\over \tau_{tr}(\mu)} = {2\pi \over \hbar} \sum_Q
(1-\cos \theta)
{|M(Q)|^2 \over \sinh(\beta\hbar\omega_Q)} \left \{
\delta(\varepsilon_{{\bf k}+{\bf Q}_\parallel} - \mu - \hbar \omega_Q)
+ \delta(\varepsilon_{{\bf k}+{\bf Q}_\parallel} - \mu + \hbar
\omega_Q) \right \}\,.
\label{eqn18}
\end{equation}
In obtaining this result, we have made use of the momentum
conservation condition $|M_{\bf kk'}(Q)|^2
= |M(Q)|^2 \delta_{{\bf k'},{\bf k}+{\bf Q}_\parallel}$
to perform the sum over ${\bf k}'$.
The magnitude of ${\bf k}$ in Eq.~(\ref{eqn18}) is now
restricted to the Fermi wavevector. A similar approximation
in Eq.~(\ref{eqn28}) would allow the average lifetime defined there
to be replaced by Eq.~(\ref{eqn18}).

\begin{table}
\caption{The various parameters for the samples. $l_e$ and
$l_i =a/T$ were obtained from the fits of the data to the
field dependence of $\sigma_{xx}$.}

\begin{tabular}{cccccc}
&$n\,(10^{16}$m$^{-2}$) & $\mu$\,(m$^2$/Vs)
        & $l_e$\,(nm) & $a\,(\mu$m\,K) & $k_F l_e$  \\
\tableline
S1 &  3.6  &  0.134 &  35  &  14 &  20    \\
S2 &  2.21  & 0.144 &  42  &  21 &  13    \\
\end{tabular}
\end{table}

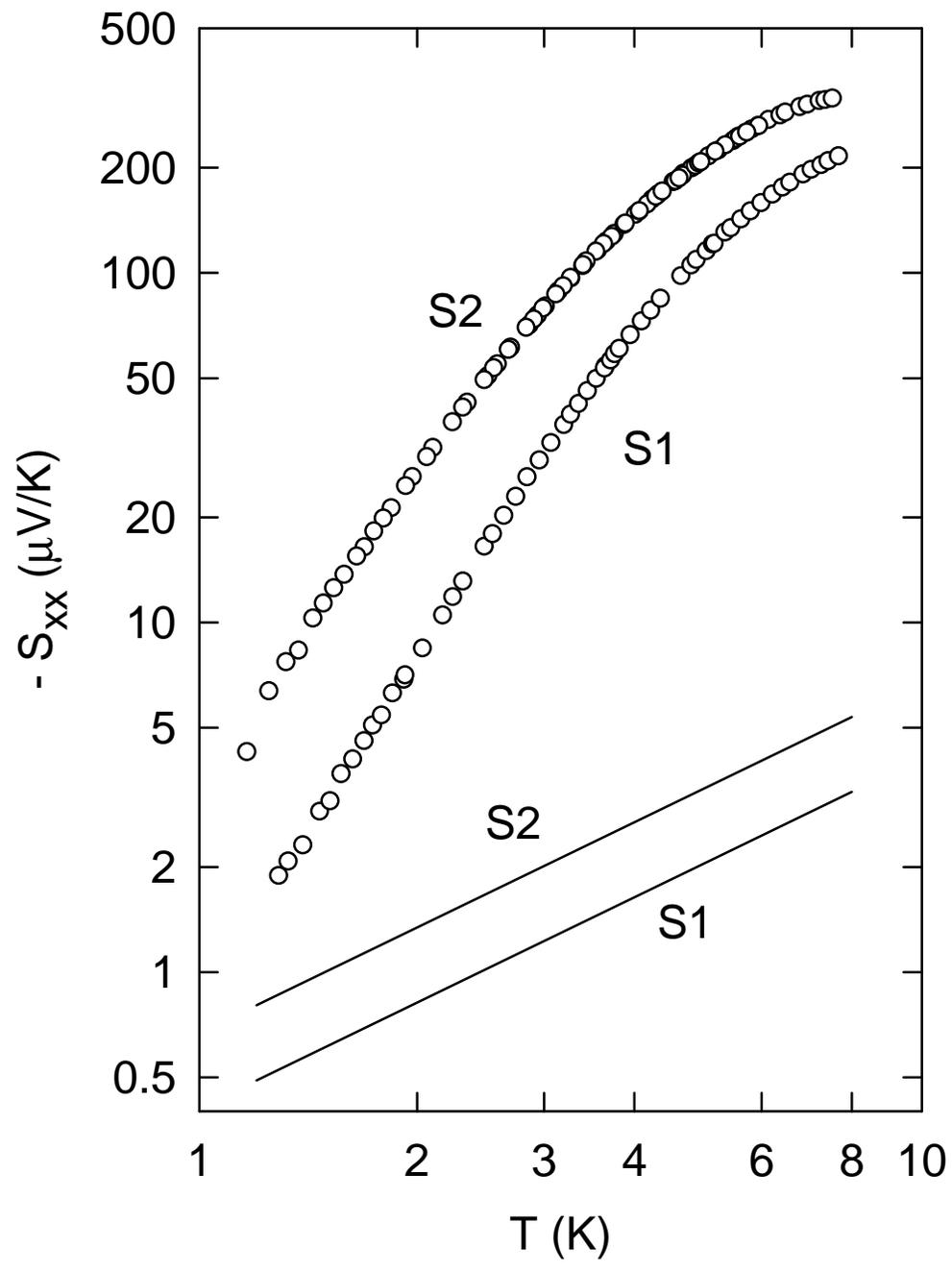
\begin{figure}
\caption{The open circles give the measured thermopower $S_{xx}$
of the two samples. The lines are the diffusion components $S^d_{xx}$.
In the case of S1, $S^d_{xx} = -0.41\,\mu$V/K is a measured quantity
taken from Ref. 13. For S2, $S^d_{xx} = -0.67\,\mu$V/K is estimated
from that of S1 assuming $S^d_{xx} \propto 1/n$. }
\end{figure}

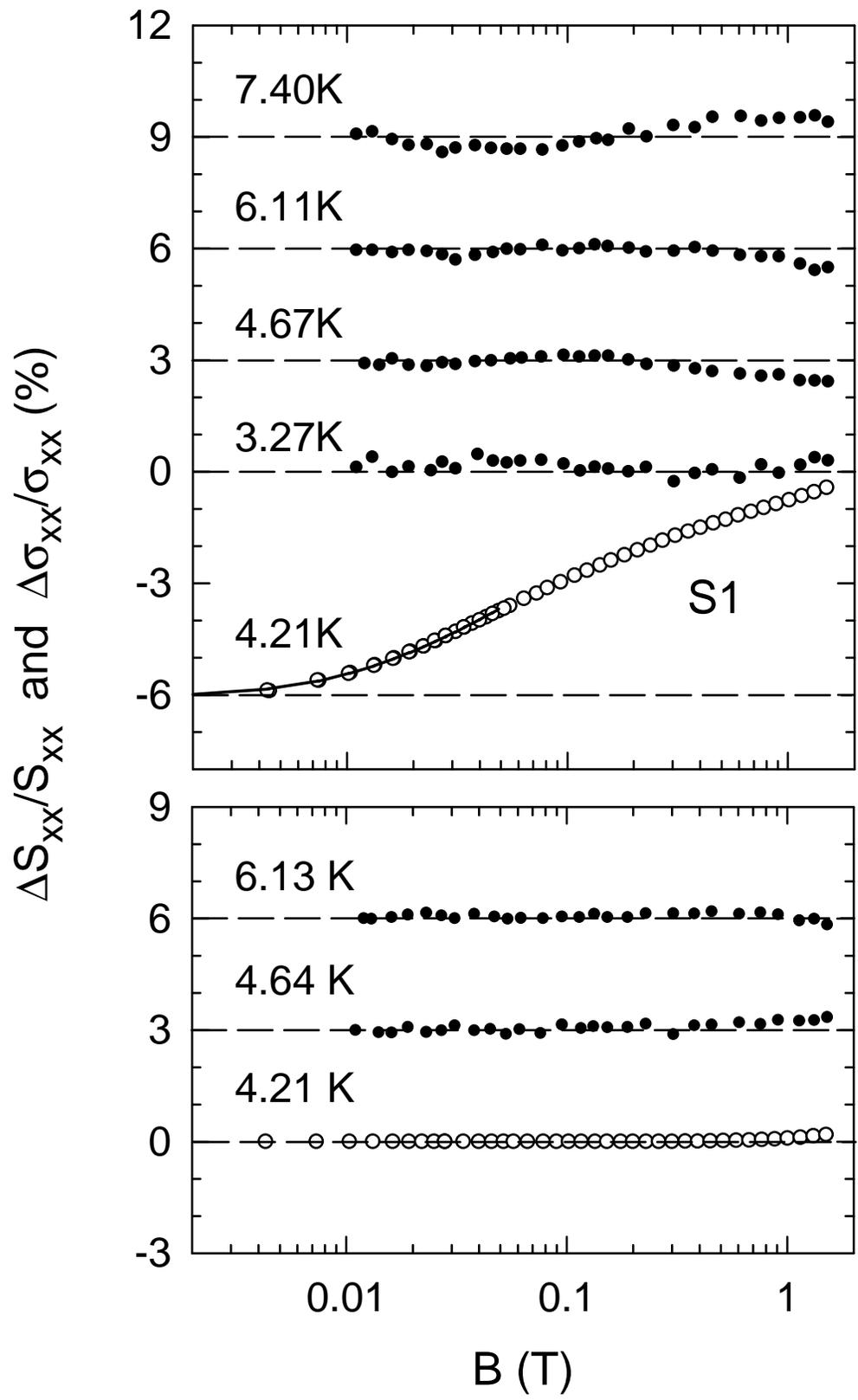
\begin{figure}
\caption{A selection of the experimental data on the field
dependence of $\sigma_{xx}$ ($\circ$) and $S_{xx}$ ($\bullet$)
at various temperatures for S1. The results are averages of $\pm B$
data, though any dependence on the direction of $B$ is weak.
The upper and lower panels
give data with $\bf B$ perpendicular and parallel to the 2DEG
respectively.  For clarity the data have been offset vertically
by multiples of 3\%. The solid line through the data for
$\sigma_{xx}$ in the upper panel is a theoretical fit according
to Eq.~(\ref{WLvsB}).}
\end{figure}

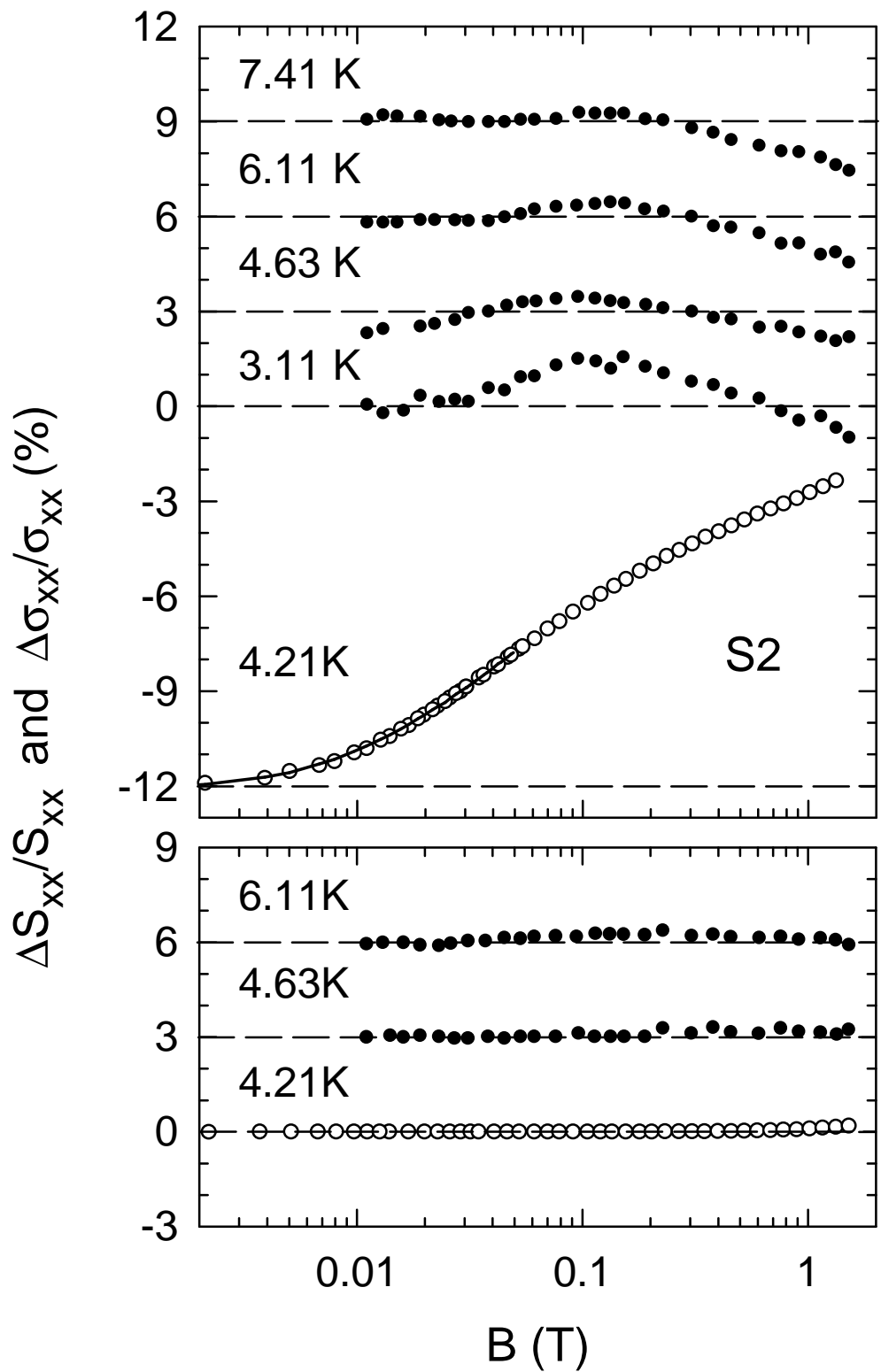
\begin{figure}
\caption{The same as Fig.~2 except these data are for S2. }
\end{figure}

\end{document}